\documentclass[final]{svjour2}
\usepackage{graphicx}
\usepackage{rotating}
\usepackage{amssymb}
\usepackage{mathptmx}
\usepackage{natbib}
\makeatletter
\journalname{Journal of Low Temperature Physics}

\newcommand{\arcmin}{\hbox{$^\prime$}}               

\newcommand{\mytilde}{\raise.17ex\hbox{$\scriptstyle\mathtt{\sim}$}} 
\newcommand  \gtsim  {\lower.5ex\hbox{$\; \buildrel > \over \sim \;$}}
\newcommand  \ltsim  {\lower.5ex\hbox{$\; \buildrel < \over \sim \;$}}

\def\3he{$^3{\rm He}$}

\def\blast{{BLAST}}

\newcommand{\wmap}{{\sl WMAP}}
\newcommand{\planck}{{\sl Planck}}

\def\lsim{\mathrel{\lower2.5pt\vbox{\lineskip=0pt\baselineskip=0pt
           \hbox{$<$}\hbox{$\sim$}}}}

\def\gsim{\mathrel{\lower2.5pt\vbox{\lineskip=0pt\baselineskip=0pt
           \hbox{$>$}\hbox{$\sim$}}}}

\bibpunct{}{}{,}{s}{}{,}

\begin{document}
\setcitestyle{numbers,square}

\newcommand{\hdblarrow}{H\makebox[0.9ex][l]{$\downdownarrows$}-}
\title{BFORE: The B-mode Foreground Experiment}

\vspace{-.5in}
\author{Michael D. Niemack$^1$ \and Peter Ade$^2$ \and
Francesco de Bernardis$^1$ \and
Francois Boulanger$^3$ \and
Sean Bryan$^4$ \and
Mark Devlin$^5$ \and
Joanna Dunkley$^6$ \and
Steve Eales$^2$ \and
Haley Gomez$^2$ \and
Chris Groppi$^4$ \and
Shawn Henderson$^1$ \and
Seth Hillbrand$^7$ \and
Johannes Hubmayr$^8$ \and
Philip Mauskopf$^4$ \and
Jeff McMahon$^9$ \and
Marc-Antoine Miville-Desch\^{e}nes$^3$ \and
Enzo Pascale$^2$ \and
Giampaolo Pisano$^2$ \and
Giles Novak$^{10}$ \and
Douglas Scott$^{11}$ \and
Juan Soler$^3$ \and
Carole Tucker$^2$}

\institute{$^1$ Department of Physics, Cornell University, Ithaca, NY, USA \\
$^2$ School of Physics and Astronomy, Cardiff University, Cardiff, CF24 3AA, U.K. \\
$^3$  Institut d'Astrophysique Spatiale, CNRS Universit\'{e} Paris-Sud, Orsay, France \\
$^4$ School of Earth and Space Exploration, Arizona State University, Tempe, AZ, USA \\
$^5$ Department of Physics and Astronomy, University of Pennsylvania, Philadelphia, PA, USA \\
$^6$ Department of Astrophysics, University of Oxford, Oxford OX1 3RH, U.K. \\
$^7$ Department of Physics and Astronomy, California State University, Sacramento, CA, USA \\
$^8$ National Institute of Standards and Technology, Boulder, CO, USA \\
$^9$ Department of Physics, University of Michigan, Ann Arbor, MI, USA \\
$^{10}$ Center for Interdisciplinary Exploration and Research in Astrophysics and Department of Physics and Astronomy, Northwestern University, Evanston, IL, USA \\
$^{11}$ Department of Physics and Astronomy, University of British Columbia, Vancouver, Canada \\
\email{niemack@cornell.edu}}

\maketitle

\begin{abstract}

The B-mode Foreground Experiment (BFORE) is a proposed NASA balloon project designed to make optimal use of the sub-orbital platform by concentrating on three dust foreground bands (270, 350, and $600\,$GHz) that complement ground-based cosmic microwave background (CMB) programs.  BFORE will survey $\sim$1/4 of the sky with 1.7 -- 3.7 arcminute resolution, enabling precise characterization of the Galactic dust that now limits constraints on inflation from CMB $B$-mode polarization measurements.  In addition, BFORE's combination of frequency coverage, large survey area, and angular resolution enables science far beyond the critical goal of measuring foregrounds.  BFORE will constrain the velocities of thousands of galaxy clusters, provide a new window on the cosmic infrared background, and probe magnetic fields in the interstellar medium.  We review the BFORE science case, timeline, and instrument design, which is based on a compact off-axis telescope coupled to $>$10,000 superconducting detectors.

\keywords{Cosmic Microwave Background, Dust, Foregrounds, Balloons, Superconducting Detectors, Kinematic Sunyaev-Zel'dovich Effect}

\end{abstract}
\vspace{-.2in}
\section{CMB $B$-mode polarization and foregrounds}
\vspace{-.1in}
Measurements of the cosmic microwave background (CMB) have played a critical role in the development of modern cosmology.  Current efforts  focus primarily on improving measurements of the CMB polarization anisotropies, which may contain the signature of gravitational waves generated during the inflationary epoch prior to 10$^{-30}$ seconds after the Big Bang.  This signature manifests as power in the `$B$-mode', or curl component, of the polarization power spectrum at degree angular scales with an amplitude proportional to the energy scale of inflation \citep{baumann/etal:2009}.  

Our ability to understand, characterize, and remove polarized Galactic foregrounds is likely to be the limiting factor in making a gravitational wave detection. The two dominant polarized foregrounds are synchrotron and thermal dust emission, which arise due to the magnetic field in the Milky Way. Observations of synchrotron emission by \wmap\ 
\cite{bennett/etal:2013} 
demonstrate that polarized synchrotron is larger than the CMB polarization anisotropies at large angular scales ($\ell<100$) at frequencies below $100\,$GHz, with a steeply falling spectral index. Above $100\,$GHz diffuse Galactic dust is the main source of polarized thermal emission. The polarized dust emissivity can be fit with a modified blackbody with $T=19.6\,$K and $\beta=1.59$, resulting in a steeply rising spectral index \citep{PlanckXXX:2015}.
This signal is due to emission from aspheric dust grains aligned with the Galactic magnetic field \citep[e.g.,][]{draine2009}.
Until recently there were few polarized measurements in the millimeter range, but in 2014--15 new full sky measurements at $353\,$GHz by the \planck\ satellite were released \citep{PlanckI:2015}.

\planck\ measured the dust polarization to be at the high end of optimistic estimates, confirming that dust is a dominant foreground for 
CMB polarization experiments  \citep{PlanckXXX:2015}. The $B$-mode signal observed by BICEP2/Keck at 150~GHz \citep{bicep2:2014} is consistent with most or all of the signal coming from dust \citep{BKP:2015}. Much is still unknown about the dust, including the composition of the grains, their size and efficiency of alignment; and about the morphology of the magnetic field that polarizes the grains and the resulting spectral dependence of the emission. Phenomena already seen in {\it Planck\/} data,
e.g., the non-zero $TB$ power spectrum and the $BB/EE\simeq0.5$ ratio \citep{PlanckXXX:2015}, already demonstrate that dust polarization is complicated. The characterization of the polarized dust emission, and the use of high frequency measurements to better remove the dust, will be critical as the CMB community attempts to detect primordial $B$-mode polarization \cite[e.g.,][]{kovetz/kamionkowski:2015}.

\begin{figure}[tbh]
\begin{center}
\vspace{-.2in}
\includegraphics[angle=0,width=4in]{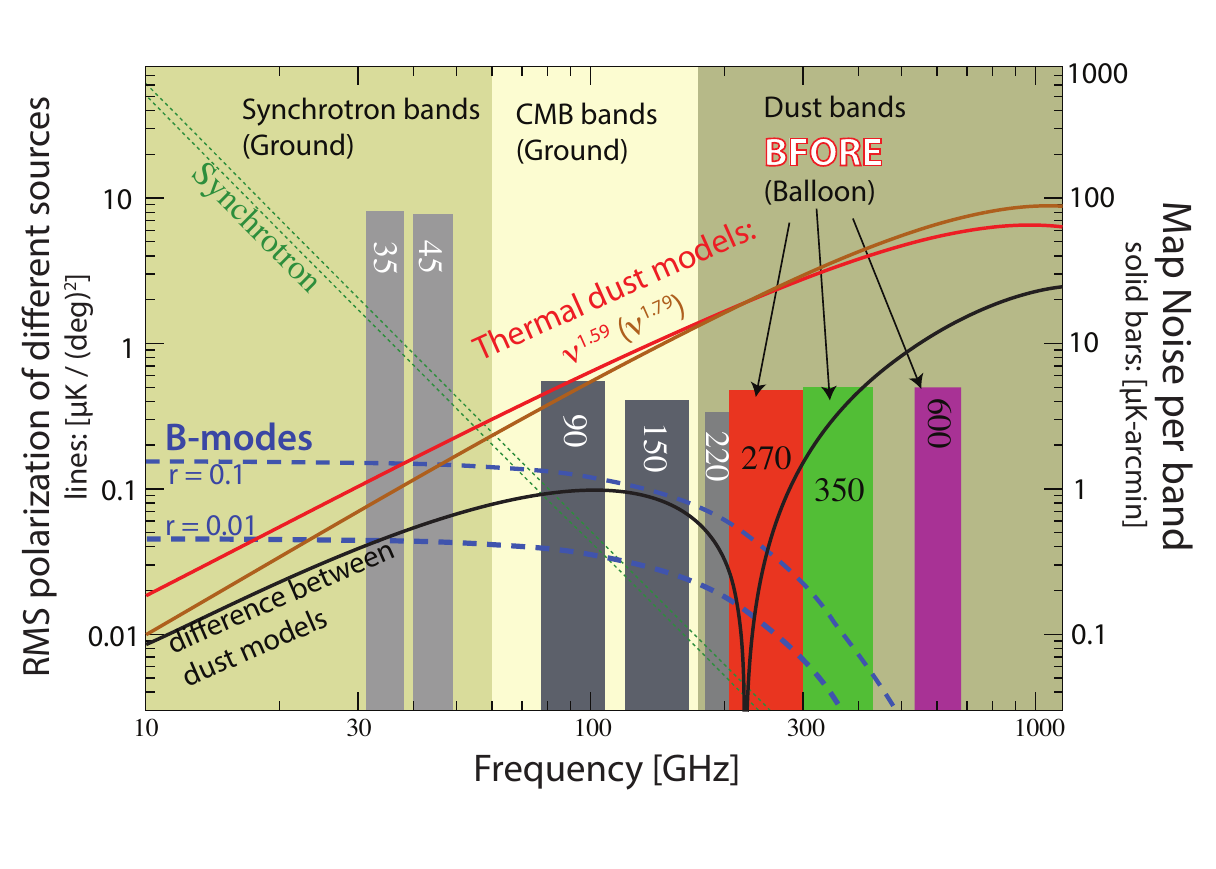}
\vspace{-.4in}
\end{center}
\caption{An estimate of the Galactic foreground polarization spectrum compared to CMB $B$-mode polarization models. The dust and synchrotron are estimated from {\it Planck} and {\it WMAP} data. The frequency bands and projected sensitivity for Advanced ACTPol \cite{henderson/etal:2015} and one BFORE flight are shown as gray and colored vertical bars, respectively.  While a sensitive ground based measurement would be limited by uncertainties in the polarized dust emission, BFORE can discriminate between competing dust models to reveal the primordial $B$-mode signal. (Color figure online)}
\label{fig:bands}
\end{figure}

\begin{figure}[tbh]
\begin{center}
\includegraphics[angle=0,width=4in]{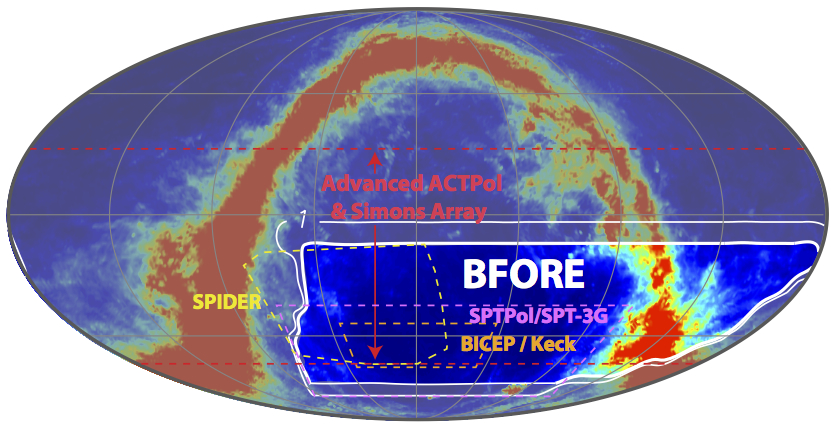}
\end{center}
\caption{Planned sky coverage for the BFORE balloon flight. This will overlap multiple complementary ground-based CMB experiments including Keck Array, SPT-3G, Advanced ACTPol, Simons Array, and the SPIDER balloon experiment \cite{filippini/etal:2010}. (Color figure online)}
\label{fig:obs}
\vspace{-.2in}
\end{figure}

The B-mode Foreground Experiment (BFORE) will utilize the sub-orbital platform by targeting three dust foreground bands (270, 350, and $600\,$GHz) that are challenging to observe from the ground due to the opacity of the atmosphere (Fig.~\ref{fig:bands}).
No other large-area higher-sensitivity observations are underway or planned in this wavelength range, with this resolution, and with a survey optimized to complement ground-based CMB observatories \cite{reichborn/etal:2010,filippini/etal:2010,lazear/etal:2014}. 
Ground-based observations are underway or planned at 220 GHz, where the dust is several times brighter than at 150 GHz and can be used as a foreground monitor, including by the 
Keck Array \cite{staniszewski/etal:2012}, Advanced ACTPol \cite{henderson/etal:2015}, Simons Array \cite{arnold/etal:2014}, and SPT-3G \cite{benson/etal:2014} projects
(Fig.~\ref{fig:obs}). However, with just one higher frequency channel, the ability to test more complex dust models and convincingly clean CMB measurements will be limited.

\begin{figure}
\begin{tabular}{cc}
\hspace{-.2in}
\begin{minipage}{2.5in}
\vspace{-.1in}
\includegraphics[height=2.in]{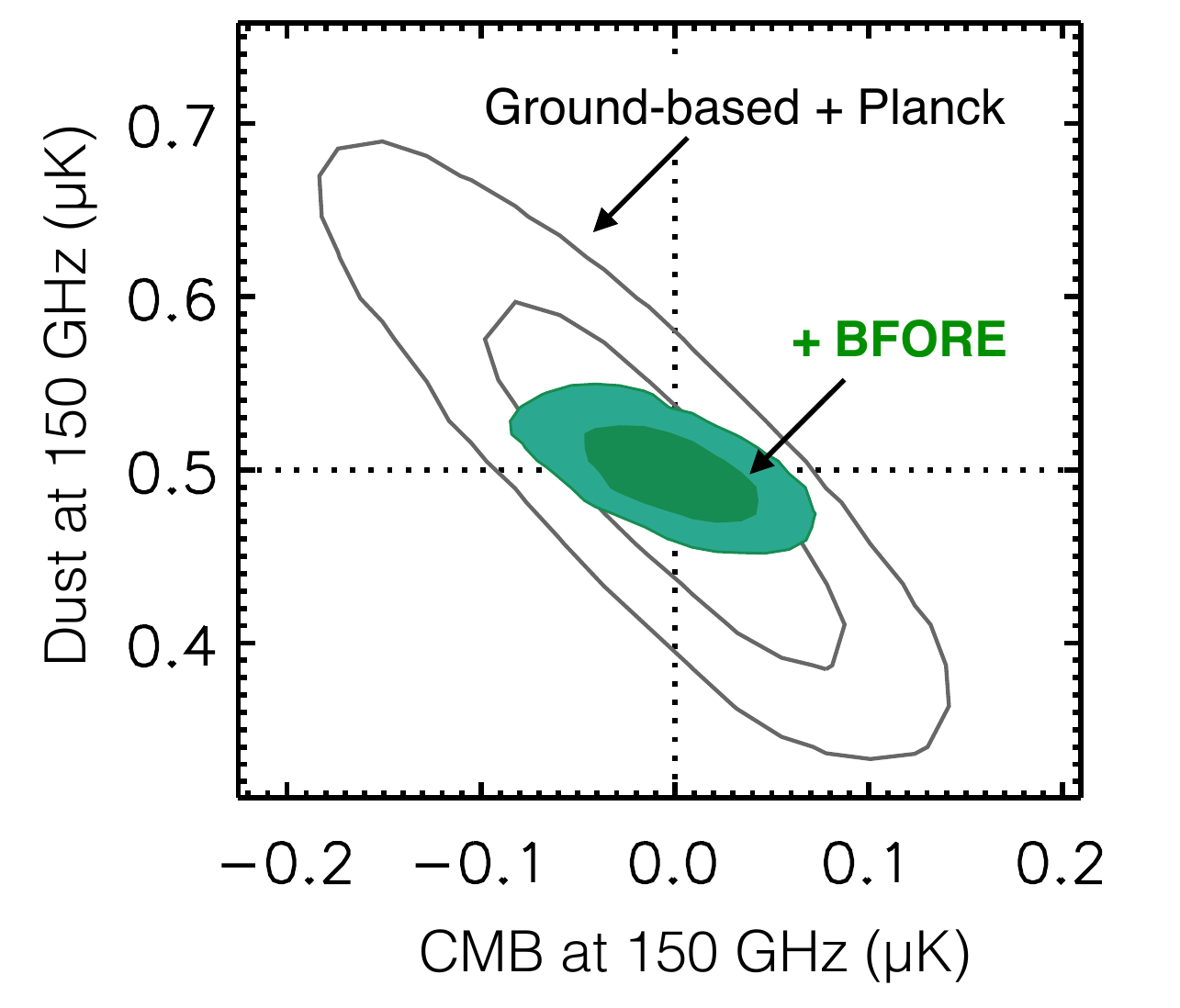}
\end{minipage}
\begin{minipage}{2.5in}
\vspace{-0.15in}
\hspace{-0.25in}
\includegraphics[height=2.in]{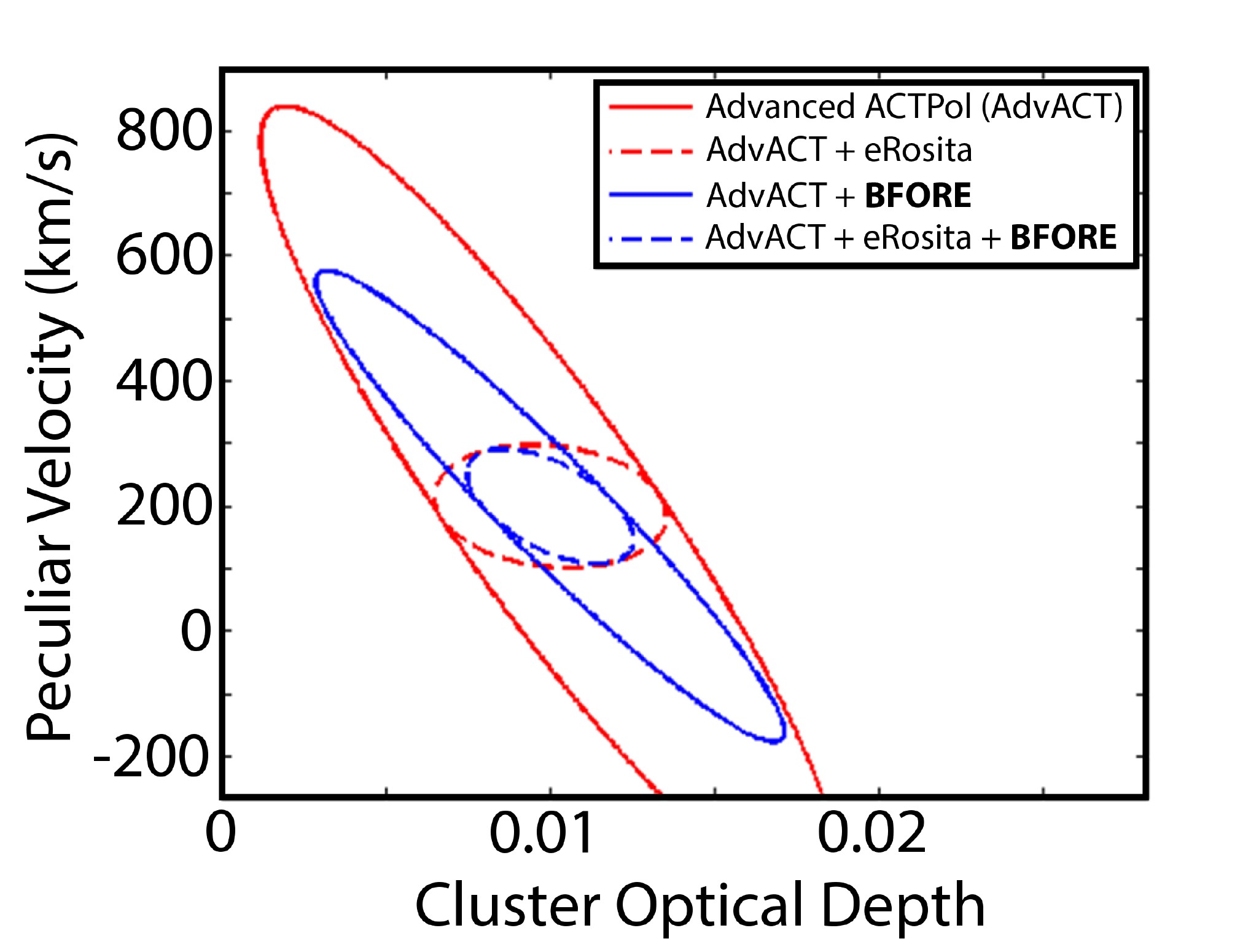}\\
\end{minipage}
\end{tabular}
\caption{\small {\bf Left:} A degeneracy exists between CMB and dust polarization at $150\,$GHz if the dust emissivity index is allowed to vary spatially (shown
here for a $25\,$deg$^2$ pixel mapped at 90--220$\,$GHz from the ground, marginalized over the index). BFORE breaks this degeneracy by measuring the dust emissivity index with a precision of $\sigma(\beta) \approx 0.03$ in hundreds of patches. {\bf Right:} Constraints on individual galaxy cluster peculiar velocities and optical depths will be substantially improved by combining BFORE with ground-based observations from Advanced ACTPol (AdvACT) \cite{henderson/etal:2015}, enabling velocity determinations of massive clusters (solid lines).  Combining this with X-ray measurements of the cluster gas temperature from {\it eROSITA}
\cite{predehl/etal:2014}
will enable velocity and optical depth measurements of a large fraction of the clusters detected by AdvACT (dashed lines). (Color figure online)}
\label{fig:kSZ}
\vspace{-.2in}
\end{figure}

To believe a result is primordial, a common spectrum signal will be required in at least two channels (typically 90 and 150\,GHz) and must be robust to changes in the foreground removal technique, foreground model, and region on the sky \citep[e.g.,][]{armitagecaplan11}. 
{There are many possible complexities in the foregrounds.} The effective emissivity index and dust temperature likely vary over the sky, and there is evidence from \planck\ that polarization fraction varies with wavelength. 
As shown in Fig.~\ref{fig:kSZ}, BFORE measurements will enable the breaking of these degeneracies.

\vspace{-.1in}
\section{Science enabled by BFORE}
\label{sec:other_science}
\vspace{-.1in}

When designing BFORE, a variety of configurations were considered. For example, if the only goal was to clean CMB $B$-mode experiments of foregrounds, a relatively small telescope aperture is sufficient to obtain resolutions in the 10\arcmin\ range.  There is, however, a wealth of other science that can be carried out in BFORE's spectral range, and this motivates the choice of a larger aperture mirror with resolution between 1$^{\prime}$ and 4$^{\prime}$.  

The Sunyaev-Zel'dovich (SZ) effect results from the scattering of CMB photons by hot ionized gas within galaxy clusters and has both thermal and kinematic components. Since blind surveys began discovering new galaxy clusters
 \cite{staniszewski/etal:2009, menanteau/etal:2010,PlanckVIII:2011}
, the thermal SZ effect has proven to be a valuable probe of large-scale structure.  The kinematic (kSZ) component was first measured by cross-correlating ACT measurements with optical measurements of luminous red galaxies \citep{hand/etal:2012}, 
but it has not yet been detected from bulk motions of individual clusters, despite targeted attempts on small numbers of massive clusters \citep[e.g.,][]{benson/etal:2003}.  Measuring the kSZ effect directly, and inferring the underlying velocity field, has great potential for improving cosmological constraints on dark energy and neutrinos
 \citep[e.g.,][]{bhattacharya2008, mueller/etal:2014}. However, separating the kSZ signals from tSZ signals, the primordial CMB, and the cosmic infrared background (CIB) is challenging and requires several frequency bands at higher sensitivity than current observations provide.

We find that combining BFORE and Advanced ACTPol measurements will substantially improve extraction of kSZ signals from individual clusters, as shown in Fig.~\ref{fig:kSZ}.  Our forecasts follow Ref.~\cite{knox/etal:2004} and account for contamination from primary CMB fluctuations, dusty star forming galaxy emission, and instrument noise. Combining BFORE and Advanced ACTPol measurements improves the kSZ figure-of-merit by $>3$ compared to Advanced ACTPol alone.

BFORE measurements will help to clean the gravitational lensing-induced $B$-mode spectrum at several arcminute scales, which offers another potential path towards improving cosmological constraints.  BFORE will complement both \planck\ and {\it Herschel} satellite measurements of the cosmic infrared background by surveying wider areas than {\it Herschel} with better sensitivity than {\it Planck}. Cross-correlation studies with CMB, optical, radio and X-ray maps will provide detailed constraints on models for the evolution of the star-formation and gas distribution within dark matter halos.  In addition, characterizing the Galactic dust will improve our understanding of the magnetic structure of our Galaxy, the nature of interstellar dust, and the turbulent energy cascade that leads to star formation \cite[e.g.,][]{hennebelle/falgarone:2012}.

\vspace{-.3in}
\section{Instrument overview}
\vspace{-.1in}
The baseline design of BFORE comprises an off-axis Gregorian telescope with a $1.35\,$m illuminated primary aperture. The  secondary mirror is cooled to $4\,$K to both reduce the loading on the detectors and provide a cold image of the primary.  The secondary is followed by a cryogenic, wide-bandwidth, stepped half-wave-plate \cite{pisano/etal:2014} and a compound field lens that focuses a $3^{\circ}\,$wide field-of-view onto a $22\,$cm focal plane cooled by a $^3$He sorption fridge.  The three observation frequencies of 270, 350, and $600\,$GHz will each have approximately 30\% bandwidth and resolutions of 3.7$^\prime$, 2.9$^\prime$, and 1.7$^\prime$, respectively.

\begin{figure}
\vspace{-0.15in}
\includegraphics[height=1.9in]{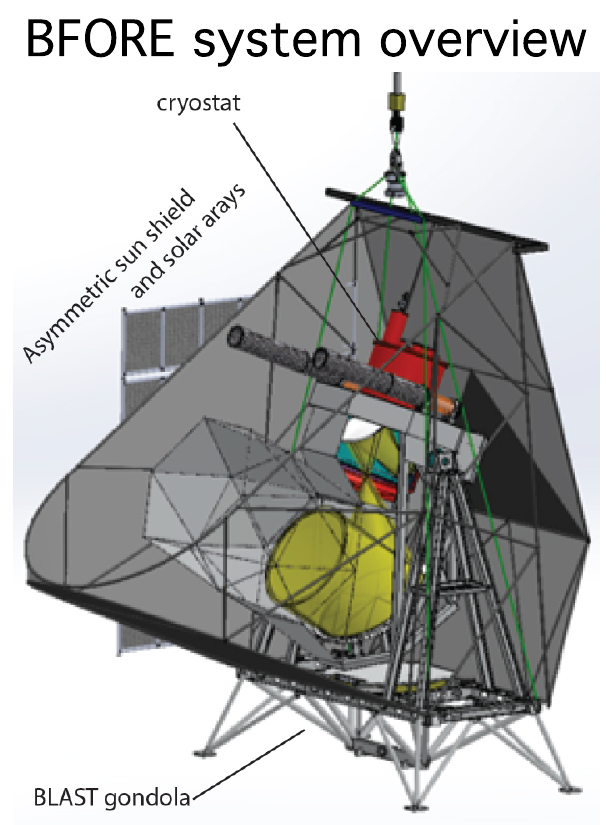}
\includegraphics[height=1.9in]{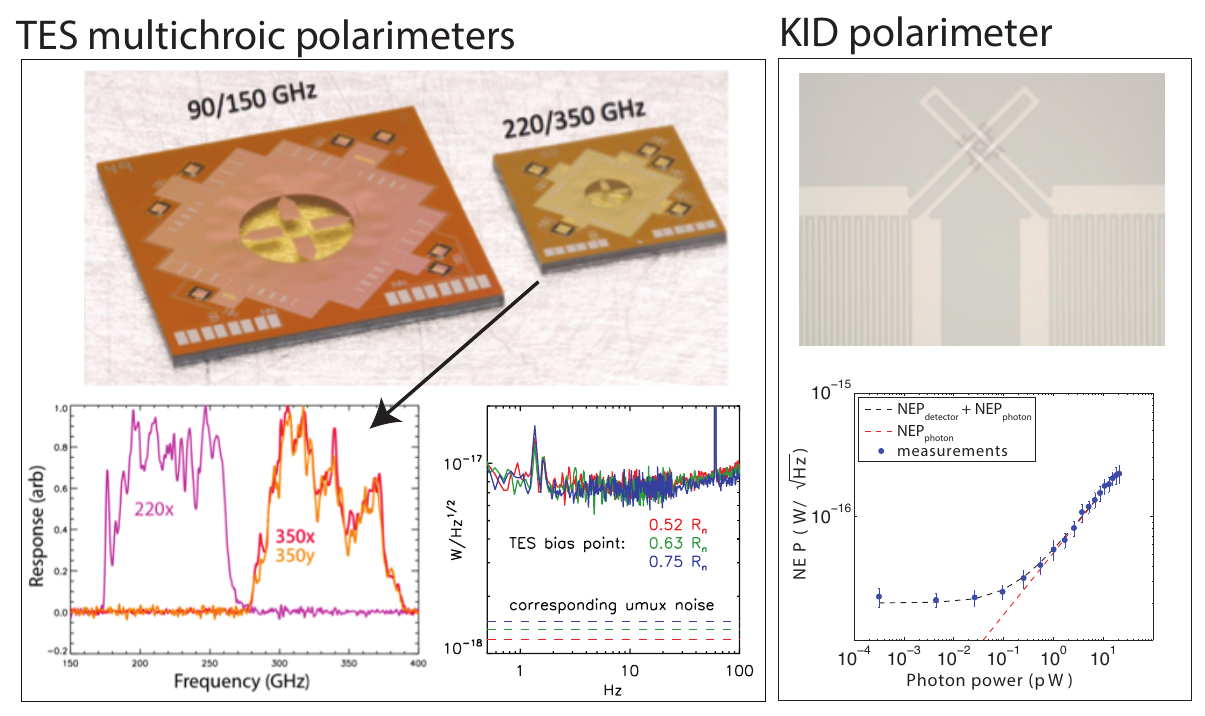}
\vspace{-0.05 in}
\caption{{\bf Left:} BFORE system overview showing the cryostat as well as the sun shields that surround the telescope mounted on the gondola. {\bf Middle:} Horn-coupled TES multichroic polarimeters for 90/150~GHz and 220/350~GHz \cite{datta/etal:2014}. Also shown are measured passbands for the 220/350 GHz polarimeters.  The noise figure shows that we have fabricated bolometers with lower NEP than is required for BFORE \citep{niemack/etal:2012}.  We also indicate the demonstrated microwave multiplexing input-referred current noise (17 pA/Hz$^{1/2}$) relative to bolometer noise measurements; the multiplexing noise is an order of magnitude below the projected BFORE detector noise. {\bf Right:} Prototype horn-coupled KID polarimeter fabricated at NIST and planned for use in BLAST.  The measured KID noise and a fit to the detector and photon noise contributions are also shown as a function of optical loading \citep{hubmayr/etal:2014}. (Color figure online)}
\label{fig:actpolprototype}
\vspace{-.2in}
\end{figure}

\vspace{-.15in}
\section{Detector arrays}
\vspace{-.1in}
The detectors for BFORE consist of arrays of roughly 2000 multichroic feedhorn-coupled polarization-sensitive detectors at 270 and $350\,$GHz, and a 2000 pixel feedhorn-coupled high frequency array at $600\,$GHz for a total of 12,000 detectors (6000 dual-polarization pixels).  
All of the detectors are packed into a single focal plane, approximately $22\,$cm in diameter, with the high frequency detectors in the center and the low frequency detectors around the outside. 
At such large detector densities, readout using standard transition-edge sensor (TES) multiplexing techniques would require complicated wiring harnesses and integration techniques, and the number of wire bonds required for a high frequency TES array would make it difficult to fit all of the detectors into the focal plane. Instead, we plan to use kinetic inductance detectors (KIDs) based on the arrays developed for \blast\ for the high frequency channel \citep{hubmayr/etal:2014} and TES detectors based on the arrays developed for ACTPol \cite{datta/etal:2014} and Advanced ACTPol \cite{henderson/etal:2015} for the low frequency channels. Photos and measurements of prototype detectors are shown in Fig.~\ref{fig:actpolprototype}. We plan to use microwave readout for the KIDs 
\cite{duan/etal:2010} 
and microwave SQUID multiplexing \citep[e.g.,][]{noroozian/etal:2015, dicker/etal:2014} for the TES arrays to achieve large multiplexing factors and maintain compatibility while minimizing cabling complexity.

\vspace{-.15in}
\section{Sensitivity forecast}
\vspace{-.1in}
After surveying 10,000$\,{\rm deg}^2$ in one 20 day Antarctic flight, the map sensitivities at 270, 350, and $600\,$GHz in Rayleigh-Jeans temperature units are all predicted to be $\sim4\,\mu$K-arcmin, or in CMB temperature units 20, 50, and $1000\,\mu$K-arcmin, respectively (polarization sensitivities are $\sqrt{2}$ higher).  For foreground removal considerations we scale these to an equivalent CMB map depth at $150\,$GHz of 3, 2, and $1\,\mu$K-arcmin, respectively.  

\vspace{-.15in}
\section{Summary}
\vspace{-.1in}
BFORE has the potential to significantly advance the search for inflationary gravity waves by characterizing CMB foregrounds many times better than \planck\ across $1/4$ of the sky and will also enable kSZ, CIB, and Galactic science. BFORE has been proposed to the NASA Astrophysics Research and Analysis Program.  
The first flight is planned for the end of 2019.

\vspace{-.1in}
\begin{acknowledgements}
The development of multichroic detectors and lenses was supported by NASA grants NNX13AE56G and NNX14AB58G.
\end{acknowledgements}

\end{document}